\documentclass{article}
\usepackage{waspaa21,amsmath,graphicx,url,times}
\usepackage{color}
\usepackage[bookmarks=false, hidelinks]{hyperref}
\usepackage{color}
\usepackage{amssymb}
\usepackage{array}
\usepackage{textcomp}
\usepackage{multirow}
\usepackage{microtype}
\usepackage{booktabs}
\usepackage{dsfont}
\usepackage{pifont}
\usepackage[pagewise,switch]{lineno}
\usepackage[dvipsnames]{xcolor}
\usepackage{bm}

\definecolor{mygray}{gray}{0.5}
\title{Self-Supervised Learning from Automatically Separated Sound Scenes}

\name{\begin{tabular}{c}Eduardo Fonseca,$^{1,2}$\sthanks{Work done during an internship at Google Research. \newline This extended version of a WASPAA 2021 submission under same title
has additional discussion for easier consumption.}
      Aren Jansen,$^{2}$
      Daniel P. W. Ellis$,^{2}$
      Scott Wisdom,$^{2}$
      Marco Tagliasacchi,$^{2}$ \\
      John R. Hershey,$^{2}$
      Manoj Plakal,$^{2}$
      Shawn Hershey,$^{2}$    
      R. Channing Moore,$^{2}$       
      Xavier Serra$^{1}$ \end{tabular}}
      
\address{$^1$Music Technology Group, Universitat Pompeu Fabra, Barcelona, {\texttt{eduardo.fonseca@upf.edu}}\\
        $^2$Google Research, New York, NY, Mountain View, CA, Cambridge, MA, and Z{\"u}rich, CH \\
}
\begin{document}

\ninept
\maketitle

\begin{sloppy}

\begin{abstract}
Real-world sound scenes consist of time-varying collections of sound sources, each generating characteristic sound events that are mixed together in audio recordings. 
The association of these constituent sound events with their mixture and each other is 
semantically constrained: the sound scene contains the union of source classes and not all classes naturally co-occur.  With this motivation, this paper explores the use of unsupervised automatic sound separation to decompose unlabeled sound scenes into multiple semantically-linked views for use in self-supervised contrastive learning.  We find that learning to associate input mixtures with their automatically separated outputs yields stronger representations than past approaches that use the mixtures alone.
Further, we discover that optimal source separation is not required for successful contrastive learning by demonstrating that a range of separation system convergence states all lead to useful and often complementary example transformations.
Our best system incorporates these unsupervised separation models into a single augmentation front-end and jointly optimizes similarity maximization and coincidence prediction objectives across the views.  
The result is an unsupervised audio representation that rivals state-of-the-art alternatives on the established shallow AudioSet classification benchmark.
\end{abstract}


\section{Introduction}
\label{sec:intro}
The construction of human-labeled audio datasets for sound event recognition is notoriously time-consuming and subjective, imposing practical limitations on dataset size and quality.
For example, label noise issues such as missing, incorrect or inconsistent labels are inherent to this process \cite{fonseca2020addressing,zhu2020audio,Fonseca2019learning,Fonseca2019model}.
In contrast, unsupervised representation learning can exploit much larger amounts of data without the need for additional labeling.
One of the advantages of these representations is they are less specialized towards the often-biased human labels and thus may be better suited for generalization to other tasks~\cite{kawakami2020learning,riviere2020unsupervised,baevski2020wav2vec,shortowards}.

Self-supervised methods aim at learning representations without the need for external supervision. 
Absent explicit labels generated by humans, the success of these methods relies on the design of \textit{proxy} learning tasks in which pseudo-labels are generated from patterns in the data.
These methods solve proxy tasks on unlabeled data to learn mappings from inputs to low-dimensional representations, which can then be used for downstream tasks such as classification.
This paradigm has seen major progress in computer vision \cite{2020_ICML_SimCLR, 2020_CVPR_MoCo,grill2020bootstrap} and in speech recognition \cite{oord2018representation,baevski2019vq,baevski2020wav2vec}.
For general-purpose audio, including a variety of environmental sounds beyond speech, the majority of works are based on contrastive learning \cite{jansen2018unsupervised,jansen2020coincidence,wang2020contrastive,saeed2020contrastive,fonseca2021unsupervised,wang2020multi}, where a representation is learned by comparing pairs of examples selected by some semantically-correlated notion of similarity~\cite{le2020contrastive}. 
Specifically, comparisons are made between positive pairs of ‘‘similar’’ and negative pairs of ‘‘dissimilar’’ examples, with the goal of learning a representation that pulls together positive pairs and thus reflects semantic structure.
One of the first works in contrastive audio representation learning uses triplet loss \cite{jansen2018unsupervised}, in which anchor-positive pairs are created by sampling neighboring audio frames as well as applying other audio transformations (e.g., adding noise).
The proxy task of \textit{coincidence prediction} learns a representation to support predicting whether a pair of examples occurs within a certain temporal proximity \cite{jansen2020coincidence}.

Recently, promising results have been attained by contrastive learning approaches that solve the proxy task of \textit{similarity maximization} \cite{saeed2020contrastive,fonseca2021unsupervised,wang2020multi}, following the seminal SimCLR work in visual representation learning \cite{2020_ICML_SimCLR}. 
This method consists of maximizing the similarity between differently-augmented \textit{views} of the same input audio example. Critical to its success is the  simultaneous use of a diversity of semantics-preserving, domain-specific augmentation methods~\cite{2020_ICML_SimCLR,fonseca2021unsupervised}.
For audio modeling, proven augmentation strategies include sampling nearby audio frames \cite{jansen2018unsupervised,jansen2020coincidence,saeed2020contrastive,fonseca2021unsupervised,wang2020multi}, artificial example mixing \cite{fonseca2021unsupervised,wang2020multi,jansen2018unsupervised}, time/frequency masking \cite{fonseca2021unsupervised,wang2020multi}, random resized cropping \cite{fonseca2021unsupervised} and time/frequency shifts \cite{jansen2018unsupervised,fonseca2021unsupervised,wang2020multi}.
In most cases, these augmentations introduce artificial, handcrafted transformations with hyperparameters that must be tuned to lie within a semantics-preserving range. 
However, typical real-world sound scene recordings already tend to be quite complex, involving mixtures of several sound sources at varying levels and unexpected channel distortions.  Therefore, these artificial augmentation techniques risk introducing an unrealistic domain shift that hinders generalization to real-world applications.

In this work, we explore using unsupervised sound separation as an alternative path to generate views for contrastive learning.  
This provides a sort of inverse to traditional example mixing augmentation: instead of constructing artificial mixtures, we decompose a naturalistic sound scene into a collection of simpler channels that share semantic commonalities with the original recording and each other.
In contrast to the previous approaches, this automatic separation approach is data-driven and input-dependent, producing ecologically valid views that eliminate the need for parameter tuning for the given dataset.  
We demonstrate that learning to associate sound mixtures with their constituent separated channels elicits semantic structure in the learned representation and is complementary to other commonly-used data augmentations when composed. 
Furthermore, we pair this augmentation procedure with a multitask objective that includes both similarity maximization and coincidence prediction, which exhibit complementary behavior for different downstream representation use cases.
Finally, we discover that a wide range of separation model competencies enable useful (and complementary) augmentations, demonstrating that optimal sound separation performance is not essential for representation learning.
The best representation learned with our sound-separation informed framework achieves an mAP of 0.326 on the downstream shallow-model AudioSet classification task.
This exceeds previous results on this benchmark under the same evaluation protocol \cite{jansen2018unsupervised,jansen2020coincidence}, and is on par with the state-of-the-art under comparable evaluation settings
\cite{wang2020multi}.

\section{Sound Separation as Data Augmentation}
\label{sec:ssep_da}
\begin{figure*}[ht]
  \vspace{-4mm}
    \centering
  \centerline{\includegraphics[width=0.65\textwidth]{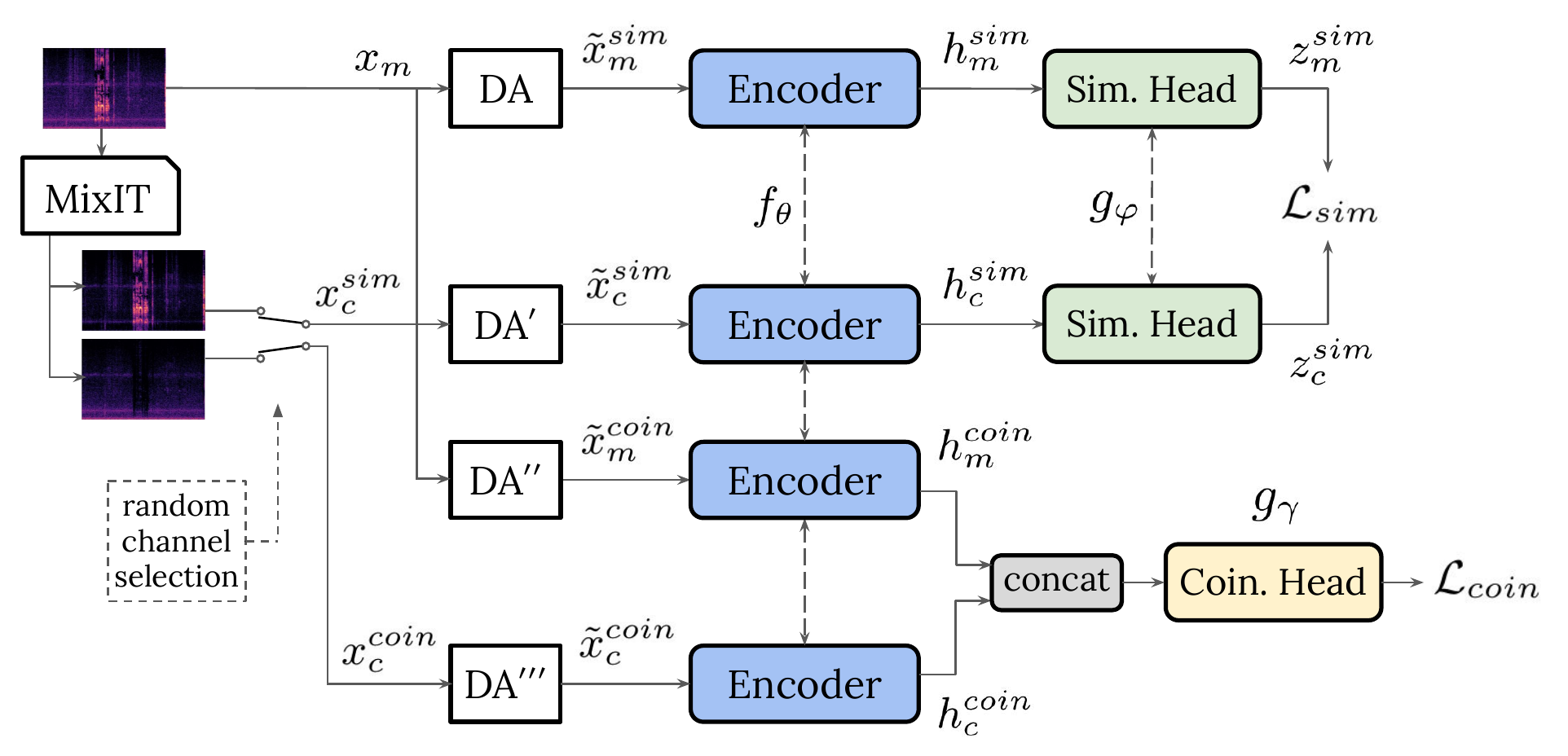}}
  \vspace{-2mm}
  \caption{\textbf{Proposed contrastive learning framework}. It is composed of an unsupervised sound separation and augmentation front-end, a common encoder $f_{\theta}$, and two task-specific heads, $g_{\varphi}$ and $g_{\gamma}$, for the similarity maximization and coincidence prediction tasks respectively. Dashed lines between networks denote shared weights. Each separated channel feeding each proxy task ($x_{c}^{sim}$ for similarity maximization or $x_{c}^{coin}$ for coincidence prediction) is selected randomly between the two output channels from the MixIT separation model. The concat block stacks the latent representations for each view to define the input to the coincidence prediction head.  Primes in the data augmentation (DA) blocks illustrate that each block is a different instance of the same augmentation policy, combining Temporal Proximity and SpecAugment. Note that the front-end illustrates the creation of pairs of positive examples---the pairs of negatives are constructed from different clips.
}
  \label{fig:diagram}
  \vspace{-2mm}
\end{figure*}

Sound separation has been studied as preprocessing to improve supervised sound event detection \cite{turpault2020improving,turpault2020sound}.
Here, we propose sound separation as an {\em augmentation} to generate pairs of positive examples for contrastive learning.
In order to come up with good views for contrastive learning, previous work argues that their mutual information must be reduced while the downstream semantically-relevant information between the views is retained \cite{2020_arXiv_GoodViews}.
Another important observation is that, since contrastive learning pulls together representations of positive views, the proxy task attempts to ignore the transformations applied to create them. Consequently, how the pairs of views are generated determines the invariant properties promoted in the learned representation.

We propose to decompose an incoming audio clip, which in general is a \textit{mixture} of multiple sound events, into its constituent sources. We then use the mixture and these separated channels to form positive pairs for contrastive learning.
In particular, the comparison of the input mixture and one of the separated channels should meet the requirements established above: \textit{(i)} the mutual information is reduced as, in principle, there is at least one input sound source that is no longer in the separated channel; \textit{(ii)} some relevant semantics are preserved as the sound source(s) present in the separated channel is also present in the input mixture. 
Therefore, in theory, this comparison would be well suited for contrastive learning. Further, with this comparison we are promoting the learning of representations that are invariant to combinations of naturally co-occurring or overlapping sources---a valuable property for general-purpose audio recognition applications.
We use this \textit{mixture vs channel} comparison as the default contrastive setup for the majority of our experiments, as illustrated in the proposed learning framework depicted in Fig.~\ref{fig:diagram}.
By contrast, the comparison between two separated channels would not be appropriate for the similarity maximization task as, in principle, each channel would contain different sources, thereby violating the semantic preservation requirement.
However, this \textit{channel vs channel} comparison could still be useful for the coincidence prediction task, where the semantic equality demand is relaxed to require only some consistency between the examples in order to support their coincidence prediction. 
We evaluate experimentally these hypotheses in Sec. \ref{sec:exps}, uncovering several nuances.

\subsection{MixIT for Unsupervised Sound Separation}
For a sound separation system we use a model trained with \textit{mixture invariant training} (\textit{MixIT}) \cite{wisdom2020unsupervised}.
MixIT is an unsupervised method in which training examples are constructed by mixing existing audio clips, and the model is tasked to separate the resulting mixtures into a number of latent sources, such that an optimal remix of the separated sources best approximates the original mixtures.
The main advantage of MixIT compared to previous methods is that it does not require knowledge of ground truth source signals, which allows leveraging large amounts of unlabeled data. 
In addition, MixIT has shown great promise in the task of universal sound separation, that is,  separating arbitrary sounds instead of specializing to (e.g.) speech \cite{kavalerov2019universal,wisdom2020FUSS}.

Our separation model was trained on AudioSet \cite{gemmeke2017audio} while ignoring all available labels and using previously proposed training settings \cite{wisdom2020unsupervised}.
The separation model used is based on an improved time-domain convolutional network (TDCN++) \cite{kavalerov2019universal}, which is similar to a Conv-TasNet \cite{luo2019conv}.
This model consists first of an encoder that maps short frames of the input waveform to a latent space. Then, separation is done in the latent space where $M$ masks are predicted for the target sources. Finally, $M$ separated waveforms are reconstructed through a decoder from the masked features. 
Separated sources are constrained to add up to the input mixture via a consistency projection layer \cite{wisdom2019differentiable}. 
Our main goal is to assess the benefits of MixIT separation pre-processing coupled with a contrastive learning back-end.

\subsection{Composition of Augmentations}
\label{ssec:composition}
Using a single transformation to generate data views has been shown as inferior to the composition of several augmentations, which is essential to obtain effective representations for vision \cite{2020_ICML_SimCLR} and audio \cite{fonseca2021unsupervised}.
By composing multiple augmentations, the goal is to define a more challenging learning task so that higher-quality representations can emerge. 
However, not all compositions are necessarily valid; rather, the elements in the composition must adequately complement each other \cite{2020_ICML_SimCLR,fonseca2021unsupervised}.

Here, in order to construct a more challenging proxy task, we combine sound separation with \textit{temporal proximity} sampling (TP) \cite{jansen2018unsupervised} and SpecAugment \cite{2019_INTERSPEECH_SpecAugment} (in this order). 
Temporal proximity sampling consists of randomly selecting two audio snippets (of 0.96s in our case) as basic units to construct pairs of examples, instead of leveraging entire AudioSet clips (of typically 10s).
When randomly sampling audio snippets within a prescribed temporal proximity, we are likely to sample \textit{(i)} the same sound sources emitting somewhat different acoustic patterns as they evolve over time; or \textit{(ii)} different sources that are related semantically or casually with the initial ones.
Thus the temporal coherence among neighboring audio snippets implies a natural form of data augmentation. 
This simple method has been proven effective in many contrastive audio representation learning works \cite{jansen2018unsupervised,jansen2020coincidence,fonseca2021unsupervised,saeed2020contrastive,wang2020multi}, analogous to the common practice of random cropping with images \cite{2020_ICML_SimCLR}.

Then, we apply SpecAugment on the log-mel spectrograms of the selected audio snippets, with time warping and time/frequency masking \cite{2019_INTERSPEECH_SpecAugment}.
SpecAugment has gained popularity as data augmentation in supervised classification, and has also been used to generate views for contrastive learning of speech \cite{2020_ICMLW_SpeechSimCLR} and audio \cite{fonseca2021unsupervised}. 
The combination of TP and SpecAugment is represented by the data augmentation (DA) blocks in Fig. \ref{fig:diagram}.
Together with the preceding unsupervised sound separation stage, they form the front-end for the two proxy tasks.
In Sec. \ref{ssec:exp_ckpts} we further extend these compositions including different convergence states of the separation model which provide distinct transformations to the incoming audio.

\newpage
\section{Proxy Learning Tasks}
\label{sec:tasks}
This section describes the two proxy tasks used in our framework: a similarity maximization task, and a coincidence prediction task.
For both, a pair of positive examples is constructed by selecting two audio snippets within the \textit{same} 10~s AudioSet clip, either from the input mixture or from the resulting separated channels. 
Analogously, a pair of negative examples is constructed by selecting two snippets from \textit{different} clips (mixtures or separated channels).
This is based on the assumption that two snippets within a given temporal proximity are much more likely to be semantically related than two snippets from independent recordings. 

\subsection{Similarity Maximization}
\label{ssec:sim_task}
The similarity maximization task consists of maximizing the agreement between differently-augmented views of the same audio example.
After deriving the different views as above, their corresponding embedding representations are compared using a contrastive loss.
This loss attempts to co-locate the two representations in the same spot of the embedding space, thus promoting invariance to the transformations applied to generate the views.
This task is based on recent work on visual representation learning, SimCLR \cite{2020_ICML_SimCLR},
and its block diagram is depicted in the top half of Fig. \ref{fig:diagram}.

\smallskip
\noindent \textbf{Front-end.} The pipeline starts with one input mixture $x_{m}$ (i.e., one AudioSet clip).
Using the separation model, every incoming mixture $x_{m}$ is separated into two output channels, from which one is randomly selected for this proxy task, $x_{c}^{sim}$.
Next, $x_{m}$ and $x_{c}^{sim}$ undergo TP and SpecAugment transformations (see Sec. \ref{ssec:composition}).
Note that each example $x_{m}$ or $x_{c}^{sim}$ undergoes a different instance of the same transformation policy (indicated by DA and DA$^{\prime}$ in Fig. \ref{fig:diagram}).

\smallskip
\noindent \textbf{Encoder Network.} The outputs from the DA blocks, $\tilde{x}_{m}^{sim}$ and $\tilde{x}_{c}^{sim}$, feed a convolutional encoder $f_{\theta}$ in order to extract low-dimensional embeddings $h$.
Specifically, for the top branch we obtain $h_{m}^{sim}=f_{\theta}\left(\tilde{x}_{m}^{sim}\right)$, where $h_{m}^{sim}$ is the representation after a $d$-dimensional embedding layer and $\theta$ are the encoder's parameters. 
Once the training is over and the encoder has converged, the representation $h$ is evaluated on downstream tasks.

\smallskip
\noindent \textbf{Similarity Head.} We use a simple MLP, $g_{\varphi}$ with parameters $\varphi$, to map the representation $h$ to the final metric embedding $z$, the domain in which the contrastive loss is applied \cite{2020_ICML_SimCLR}.
This head is used to allow the representation $h$ to back away from the representation at the training objective. 
Previous work reports better downstream performance using $h$ instead of $z$ \cite{2020_ICML_SimCLR}, something we also observed in preliminary experiments.

\smallskip
\noindent \textbf{Contrastive Loss.} To compare a positive pair of examples, $x_{m}$ and $x_{c}^{sim}$, we adopt the normalized temperature-scaled cross-entropy (\textit{NT-Xent}) loss \cite{2020_ICML_SimCLR,le2020contrastive} given by
\begin{equation}
\mathcal L_{sim_{i,j}}=-\log\frac{\exp\left(\textrm{sim}(z_{i},z_{j})/\tau\right)}{\sum_{v=1}^{2N}\mathds{1}_{v\neq i}\exp\left(\textrm{sim}(z_{i},z_{v})/\tau\right)}\label{eq:NTXent}
\end{equation}
where $z_{i}$ and $z_{j}$ are the metric embeddings corresponding to $x_{m}$ and $x_{c}^{sim}$, $\textrm{sim}(\bm{u},\bm{v})= \bm{u}^\top\bm{v}/\|\bm{u}\|\|\bm{v}\|$ represents cosine similarity whose sensitivity is adjusted by a temperature value $\tau\in\left(0,1\right]$, $\mathds{1}_{v\neq i}\in\left\{ 0,1\right\}$ is a function that returns 1 when $v\neq i$, and $N$ is the batch size.
Since two views are generated from each incoming audio clip, the batch size is extended from $N$ to $2N$ elements, allowing for one pair of positive examples and $2(2N-2)$ pairs of negative examples for every input mixture---the overall loss includes both $\mathcal L_{sim_{i,j}}$ and $\mathcal L_{sim_{j,i}}$.
By minimizing the objective in (\ref{eq:NTXent}), parameters $\theta$ and $\varphi$ are adjusted to maximize the numerator (i.e., the agreement between embeddings of positives, assigning them to neighboring representations) while simultaneously minimizing the denominator (i.e., the similarity between embeddings of negatives, forcing them to distant spots in the embedding space).

\subsection{Coincidence Prediction}
\label{ssec:CP_task}

The coincidence prediction task relies on the \textit{slowness prior} in representation learning \cite{wiskott2002slow}.
Audio waveforms of sound sources can vary quickly, whereas the corresponding perceived semantics change at a much slower rate. 
Consequently, there should be a relatively stable latent representation in order to explain the semantic perception.
This representation would support the prediction of whether a pair of examples are coinciding within a given temporal proximity.
Note this task is a generalization of the correspondence prediction task proposed for audio-visual multimodal learning \cite{arandjelovic2017look}, where the task is to predict time \textit{correspondence} between audio and video frames. 
Here, we relax the time scale requirement to predict \textit{coincidence} within a prescribed temporal proximity, specifically within the (maximum) 10~s of AudioSet clips.
The task diagram is depicted in the bottom half of Fig. \ref{fig:diagram}.

\smallskip
\noindent \textbf{Front-end and Encoder network.}
In previous work, this proxy task has been applied directly to audio snippets drawn from the same/different temporal proximity \cite{jansen2020coincidence}.
Here, we adopt the augmentation front-end described in Sec. \ref{sec:ssep_da}, which is the same as for the similarity maximization task, leading to the augmented examples $\tilde{x}_{m}^{coin}$ and $\tilde{x}_{c}^{coin}$ from $x_{m}$ and $x_{c}^{coin}$.
Then, we use a convolutional encoder to extract $d$-dimensional embedding representations $h$---the same encoder network is shared across both proxy tasks, $f_{\theta}$.

\smallskip
\noindent \textbf{Coincidence Head.}
Once the embedding representations for a pair of examples are obtained, we use a coincidence network, $g_{\gamma}$ with parameters $\gamma$, tasked to predict the (non)-coincidence between the pair.
More specifically, we feed $g_{\gamma}$ with the concatenation of the two embeddings $[h_{m}^{coin},h_{c}^{coin}] \in \mathbb{R}^{2d}$.
The coincidence head consists of an MLP with one output unit, mapping the concatenated embedding representation to the probability that the input pair is coinciding---a binary classification task.

\smallskip
\noindent \textbf{Loss.}
In a generic batch of $N$ pairs of within-clip coinciding examples (i.e., positive pairs), $X=\left\{(x_1^i,x_2^i)\right\}_{i=1}^N$, we define $N-1$ pairs of negative examples per each pair of positives.
This is done by pairing the non-coinciding examples $(x_1^i,x_2^j)$ for $i\neq j$.
In this setting, for a given batch $X$ and focusing on our goal of optimizing the representation $h$, the coincidence loss function follows the class-balanced binary cross entropy expression \cite{jansen2020coincidence}:
\begin{equation}
\begin{aligned}
\mathcal L_{coin}&(X) = -\frac{1}{N}\sum_{i=1}^{N}\log g_{\gamma}([h_{m}^{coin,i},h_{c}^{coin,i}]) \\
& - \frac{1}{N(N\!-\!1)}\sum_{\substack{1\leq i,j \leq N \\ j \neq i}}\log\Big[1-g_{\gamma}([h_{m}^{coin,i},h_{c}^{coin,j}])\Big].
\end{aligned}
\label{eq:bce_coin}
\end{equation}


\subsection{Joint Optimization}
\label{ssec:discussion_tasks}
We conjecture that jointly optimizing the two objectives above in a multi-task setting can favor learning complementary information for semantic representation learning.
Both proxy tasks share the ultimate goal of contrastive learning, that is, supporting relationships between pairs of positives and pairs of negatives so as to force a semantically structured embedding space.
However, each task pursues this goal in a slightly different way, in terms of underlying principle and implementation.

\smallskip
\noindent \textbf{Underlying principle.}
The similarity maximization task essentially aims to co-locate the representations of both examples in a positive pair at the same point in the embedding space.
Therefore, for successful representation learning, it is usually required that some semantic relationship is preserved between the two examples, e.g., the two examples share some class label(s).
By contrast, the coincidence prediction task is based on a weaker condition.
Instead of co-locating representations, the goal is to assign a representation that supports coincidence prediction, establishing a clear relationship between the representations for both examples, but not necessarily requiring their collocation.

\smallskip
\noindent \textbf{Implementation.}
The NT-Xent loss of (\ref{eq:NTXent}) follows a canonical version of contrastive loss, explicitly measuring similarity of embeddings as the scoring function \cite{le2020contrastive}.
By contrast, the binary cross entropy loss of (\ref{eq:bce_coin}) is not a contrastive loss per se, but rather a loss typically used for classification, fed with probabilities.
In this case, one could argue that the coincidence head serves as a learned similarity measure between two points in the embedding space, conceptually analogous to the handcrafted scoring functions typically present in the canonical contrastive losses (e.g., cosine similarity in the NT-Xent loss).

\section{Experimental Setup}
\label{sec:setup}



\subsection{Evaluation Methodology}
\label{ssec:eval}

By optimizing the training objectives of (\ref{eq:NTXent}) and (\ref{eq:bce_coin}), the goal is to learn semantically discriminative audio representations.
We train our framework using a superset of the AudioSet training set consisting of around 3M audio clips, while ignoring all available labels.
To evaluate the learned representation $h$ we use the trained encoder $f_{\theta}$ as a feature extractor for the two evaluation methodologies considered in past work~\cite{jansen2018unsupervised,jansen2020coincidence}:

\smallskip
\noindent \textbf{Query by Example Retrieval (QbE).}
Given a small subset of AudioSet with around 100 examples per class, cosine distance is computed between all the within-class target pairs, and all (present, not-present) pairs as non-target trials.
Then, we sort the resulting distances in ascending order and compute per-class average precision (AP) of ranking target over non-target trials.
Averaging per-class AP leads to the reported \textit{QbE} mAP. 
This is a direct measurement of the representation semantic consistency without requiring further training.

\smallskip
\noindent \textbf{Downstream Classification with Shallow Model.}
This is a supervised classification task carried out by training and evaluating a shallow architecture on top of the fixed embeddings previously learned.
In particular, we use an MLP with one 512-unit hidden layer and ReLU activation, followed by a 527-way classification layer with sigmoid activation.
For this purpose, we use the entire AudioSet training set version and report \textit{classification} mAP.
This measures the usefulness of the learned representation for a large-vocabulary downstream supervised classification task.

For every experiment, we train our framework until QbE convergence, which typically occurs between 400k and 600k steps, after which QbE mAP plateaus. 
We select an encoder checkpoint from this plateau and report the QbE mAP.
Then, we use this checkpoint to extract features for the entire AudioSet and conduct the shallow classifier evaluation. After L2-normalizing the embeddings, we train on the AudioSet training set, allowing 5\% for validation where we optimize mAP, then report classification mAP on the evaluation set.

\subsection{Implementation and Training Details}
\label{ssec:imp_details}

A critical parameter in the proposed framework is the number of output waveforms in the separation model $M$, which must be defined in advance.
After experimenting with $M=\left\{ 2,4\right\}$, we decided to use $M=2$, as results with $M=4$ were slightly worse for both proxy tasks.
We attribute this to the fact that when $M=4$ it is common to produce near-empty output channels, which is problematic for the creation of positive pairs.
(This issue is further explained in Appendix \ref{sec:selectM}.)
Note that sound separation is only used during the learning of the representation---in our downstream tasks no separation is applied.
The DA blocks in Fig. \ref{fig:diagram} consist first of Temporal Proximity sampling, i.e., random selection of 0.96~s waveform snippets within the (maximum) 10~s AudioSet clips.
Snippets are transformed to log-mel spectrogram patches using a 25~ms Hann window with 10~ms hop, and 64 mel log-energy bands, leading to time-frequency patches of $T \times F=96 \times 64$. 
SpecAugment is then applied using \textit{(i)} two frequency masks and two time masks, with a max width of 10 bands or frames, respectively; and \textit{(ii)} time warping with 8 frames as maximum warp \cite{2019_INTERSPEECH_SpecAugment}.

For the encoder we use a convolutional network based on CNN14 from previous work \cite{kong2019panns}.
Our modifications from the original CNN14 include removing Batch Normalization \cite{ioffe2015batch} and Dropout \cite{srivastava2014dropout}, which was not found to be beneficial in our experiments.
In addition, we substitute the classifier layer and the preceding fully-connected layer by an embedding convolutional layer with \textit{d} filters, followed by a global max pooling operation to produce the \textit{d}-dimensional representation $h$, which is used for downstream tasks.
We use $d=128$ unless stated otherwise.
The resulting encoder network has 76M weights.
The similarity head consists of an MLP with one hidden layer of 256 units and ReLU non-linearity, followed by an output layer with 128 units, which is the dimension for the metric embeddings $z$ feeding the NT-Xent loss.
The coincidence head consists of an MLP with two hidden layers of 512 units and ReLU nonlinearities, followed by an output layer with one single unit to produce coincidence predictions feeding the class-balanced binary cross entropy loss.

Experiments are carried out considering each proxy task individually as well as the full framework trained jointly. 
When both tasks are trained jointly, the two objectives are optimized from scratch and equally weighted obtaining a joint loss $\mathcal L_{joint} = \mathcal L_{sim} + \mathcal L_{coin}$,
using one optimizer to update all the networks.
We use the Adam optimizer \cite{kingma2014adam} with a learning rate of 1e-4 whenever the coincidence prediction task is involved, or 3e-4 when only the similarity maximization task is considered.
The temperature parameter in (\ref{eq:NTXent}) is set to $\tau=0.3$.
Learning rates and $\tau$ are tuned by optimizing QbE mAP on a validation set different from that used to report results.

The framework is trained on Google Cloud TPUs of 32 cores with a global batch size of 2048, which means local batches of 64 examples per core.
Loss contributions and gradients are computed locally in each replica, then aggregated across replicas before applying the gradient update.
Contrastive learning approaches typically benefit from comparison with multiple negative examples.
In our framework, negative examples are drawn from clips within the current batch at every iteration, as done in previous works \cite{2020_ICML_SimCLR,fonseca2021unsupervised,jansen2020coincidence,wang2020multi}.
This approach is more practical than relying on a memory bank \cite{2018_CVPR_InstanceDis}, a memory queue \cite{he2020momentum}, or negative mining techniques to find suitable negatives \cite{jansen2018unsupervised}.
However, with this simple approach the quality and diversity of negatives are limited by the batch size (in our case, the local batch size).
Recent works show how increasing batch sizes provide solid improvements in visual \cite{2020_ICML_SimCLR} and audio \cite{wang2020multi} contrastive representation learning, the latter work utilizing batch sizes of up to 32k examples.
Here, we do not explore this 
avenue and evaluate our proposed approach using a more usual batch size. 
Based on previous literature \cite{2020_ICML_SimCLR,wang2020multi,le2020contrastive}, if our approach shows promise using the batch size selected for our experiments, it is expected to provide better performance under more favorable conditions given by larger batches.

\section{Experiments}
\label{sec:exps}

This Section describes the experiments run using the framework of Fig. \ref{fig:diagram}, or portions of it.
For simplicity, in the following Tables the similarity maximization task and the coincidence prediction task are sometimes referred to as \textit{SimCLR} and \textit{CP}, respectively.

\subsection{Baseline Experiments}
\label{ssec:exp_baseline}
Table \ref{tab:baseline} lists the performance when sound separation is ablated from the front-end in Fig. \ref{fig:diagram}, which is equivalent to all DA blocks being fed by the input mixture $x_m$.
\begin{table}[!t]
\vspace{-2mm}
\caption{mAP without sound separation in the front-end (i.e., using only the input mixture). SA = SpecAugment, TP = Temporal Proximity, CP = Coincidence Prediction.}
\vspace{+1mm}
\centering
\begin{tabular}{l|cc}
\toprule
\textbf{Representation} & \textbf{QbE mAP} & \textbf{Classif. mAP} \\
\midrule
\midrule
Log-Mel Spectrogram (baseline)                & 0.423      & 0.065 \\
\midrule
simCLR \& SA              & 0.551	    & 0.196    \\
simCLR \& TP              & 0.591	    & 0.248    \\
simCLR \& TP \& SA        & \textbf{0.613}	    & 0.265    \\
\midrule
CP \& TP \& SA   & 0.599    & \textbf{0.286}   \\
\bottomrule
\end{tabular}
\label{tab:baseline}
\vspace{-4mm}
\end{table}
We use log mel spectrogram as a baseline handcrafted representation.
As expected, both SpecAugment and TP with the similarity maximization task as back-end substantially outperform the naive mel spectrogram.
The effectiveness of TP is noteworthy considering its simplicity; 
Initially proposed in previous work \cite{jansen2018unsupervised}, it has been widely adopted in contrastive learning works \cite{jansen2020coincidence,saeed2020contrastive,fonseca2021unsupervised,wang2020multi}, some of which use it as the sole augmentation \cite{jansen2020coincidence,saeed2020contrastive}.
Note that the approach simCLR \& TP is conceptually comparable to the recent COLA \cite{saeed2020contrastive}.
Combining TP and SpecAugment outperforms either one alone, thus validating the composition in the DA blocks of the front-end.
Finally, results indicate different tendencies for the two proxy tasks, with the similarity maximization providing better QbE mAP, and the coincidence prediction attaining better classification mAP.

\subsection{Sound Separation for Contrastive Representation Learning}
\label{ssec:exp_mixit}

We now report the experiments including the unsupervised sound separation block in the front-end, as depicted in Fig.~\ref{fig:diagram}.
We assess various comparisons enabled by sound separation preprocessing, namely: \textit{(i)} comparing the input mixture with one of the separated channels (\textit{mix vs chan}); \textit{(ii)} comparing the two separated channels (\textit{chan vs chan}); or \textit{(iii)} comparing the input mixture with anything else, i.e., either with the input mixture or with one of the separated channels (\textit{mix vs any}).
Table \ref{tab:task_sim} shows the performance with the similarity maximization (\textit{SimCLR}) task as back-end.
\begin{table}[!t]
\vspace{2mm}
\caption{mAP using sound separation (SSep) in the front-end and the \textit{SimCLR} back-end. TP is always applied; SpecAugment (SA) is applied as specified.}
\vspace{+1mm}
\centering
\begin{tabular}{@{}l@{}cc|cc@{}}
\toprule
\textbf{Comparison}    & \textbf{SSep}    & \textbf{SA}    & \textbf{QbE mAP} & \textbf{Classif. mAP} \\
\midrule
\midrule
Mix vs mix (baseline)     & -       & -        & 0.591	        & 0.248   \\
Mix vs mix (baseline) & -  & \checkmark    & 0.613	        & 0.265    \\
\midrule
Mix vs chan       & \checkmark    & -     & 0.631	    & 0.272  \\
Mix vs chan       & \checkmark   & \checkmark  & \textbf{0.640} & \textbf{0.282}    \\
Mix vs any      & \checkmark   & \checkmark   & 0.638	    & 0.279    \\
Chan vs chan       & \checkmark   & \checkmark   & 0.611	    & 0.254    \\
\bottomrule
\end{tabular}
\label{tab:task_sim}
\vspace{-4mm}
\end{table}
By looking at the first rows of Table \ref{tab:task_sim}, we can benchmark SpecAugment and sound separation.
We see that sound separation preprocessing (third row) provides a bigger boost in both metrics compared with SpecAugment (second row), yet the best performance is obtained from their composition (fourth row).
This trend for the \textit{mix vs chan} comparison also holds for the other contrastive setups.

Results indicate that comparing the input mixture with the separated channels provides substantially better representations than the baseline approach of leveraging only the input mixture.
This confirms the usefulness of sound separation preprocessing for contrastive learning of audio representations. 
Allowing the input mixture to be compared with itself in addition to the separated channels (\textit{mix vs any}) does not lead to performance boosts.
Generally, the performance of \textit{mix vs any} and \textit{mix vs chan} were very similar across the experiments we ran.
Hence, we adopt \textit{mix vs chan} as best setup in order to focus on the effect of sound separation.
Finally, comparing both separated channels (\textit{chan vs chan}) performs significantly worse, on par with the non-separated baseline (for QbE mAP) or even worse (for classification mAP).
If we assume the separation has successfully isolated independent sources in each output, this comparison violates the semantic preservation principle (thus hindering the learning of semantic representations), so we might have expected decrements even larger than the $\approx$0.03 mAP with respect to \textit{mix vs chan} for both metrics.
After inspection of a few dozen separation examples, we identify two potential explanations for this observation:
First, the result of the separation algorithm is not always perfect.
This depends on the complexity of the input mixture---this is to be expected considering the great diversity of AudioSet clips.
When this happens, the same source can be present in both separated channels.
Second, even when the separation is satisfactory, there are some classes that retain a semantic relationship, e.g., two different instruments from the same family, or two different vocalizations from the same or similar animals. 
When used as a pair of positives, their relationship may still provide a useful learning signal compared to pairs of unrelated negative examples.

Table \ref{tab:task_cp} shows the performance with the coincidence prediction (\textit{CP}) task as back-end.
\begin{table}[!t]
\vspace{-2mm}
\caption{mAP using sound separation in the front-end and the CP back-end. TP and SpecAugment are applied.}
\vspace{+1mm}
\centering
\begin{tabular}{l|cc}
\toprule
\textbf{Comparison} & \textbf{QbE mAP} & \textbf{Classif. mAP} \\
\midrule
\midrule
Mix vs mix (baseline)     & 0.599    & 0.286    \\
\midrule
Mix vs chan     & \textbf{0.619} & \textbf{0.293}    \\
Chan vs chan     & 0.590	    & 0.283    \\
\bottomrule
\end{tabular}
\label{tab:task_cp}
\vspace{-4mm}
\end{table}
Similar to the \textit{SimCLR} back-end (Table \ref{tab:task_sim}), we again observe that the \textit{mix vs chan} comparison yields top performance, outperforming the no-separation baseline. Comparing both separated channels (\textit{chan vs chan}) again leads to the worst results, in this case underperforming the baseline for QbE mAP, while being on par in terms of classification mAP.
In addition to corroborating the utility of sound separation in the front-end, these results also demonstrate that \textit{CP} benefits from composing augmentations, which was not explored in previous work \cite{jansen2020coincidence}, where only TP is used.

Comparing performance across both proxy tasks, we notice that \textit{SimCLR} always yields the best QbE mAP while \textit{CP} produces top classification mAP.
This could be due to a better alignment between \textit{SimCLR}'s underlying principle (maximizing/minimizing the cosine similarity between positives/negatives) and the QbE retrieval mAP (computed by ranking pairwise cosine distances).
Finally, regarding the performance using the \textit{chan vs chan} comparison, \textit{CP} shows significantly better classification mAP than similarity maximization (specifically, a mAP on par with the latter's best case).
This accords with our intuition that \textit{CP} is tolerant of semantic differences between positives due to its weaker assumptions. 
However, for QbE mAP, the opposite behaviour is observed, presumably because this tolerance does not help the QbE objective.


\subsection{Separation Processing at Different Convergence States}
\label{ssec:exp_ckpts}

In the previous Section we show that sound separation is beneficial for our tasks, even when the separation is less than perfect as can occur when input mixtures are difficult to separate.
This leads us to ask whether the processing provided by a separation model \textit{before} convergence can also be a valid form of augmentation for contrastive representation learning. 
To answer this, we experiment with separation examples generated by multiple training checkpoints of a single separation network.
We view the separation checkpoints as \textit{audio processors} that implement complex modifications on the incoming audio.
A qualitative assessment of output streams as learning progresses indicates four types of processors corresponding to four convergence states, and we empirically characterize their behavior as follows.\footnote{Note that the description of every \textit{processor} is approximate and somewhat dependent on the input’s complexity. For example the S1 model could provide a good separation when fed with an easy mixture.}
(Fig. \ref{fig:four_ckpts} shows example spectrograms of the separated channels for each of these processors given the same input mixture.)
\begin{itemize}
    \item \textbf{Separation after full convergence (S2, 1.7M steps)}. This is the separation model used for experiments in Sec. \ref{ssec:exp_mixit}.
    \item \textbf{Separation before convergence (S1, 5k steps)}. Separation performance is more limited.
    \item \textbf{Filtering with early training model (F, 500 steps)}. Outputs are produced by the separation model very early during training. After $\sim$500 steps, sources are not separated, and the output channels are differently filtered versions of the input. Most sources in the input are present in all outputs, but often with different levels/spectral content, such as different spectro-temporal modulations.
    \item \textbf{Noise with untrained model (N, 0 step)}. Outputs are produced by the separation model untrained. They feature a clearly audible, wideband structured noise, correlated with the input signal. Audio artifacts are sometimes present. Both output channels are very similar.
\end{itemize}
\begin{figure*}[ht]
  \vspace{-4mm}
    \centering
  \centerline{\includegraphics[width=1.0\textwidth]{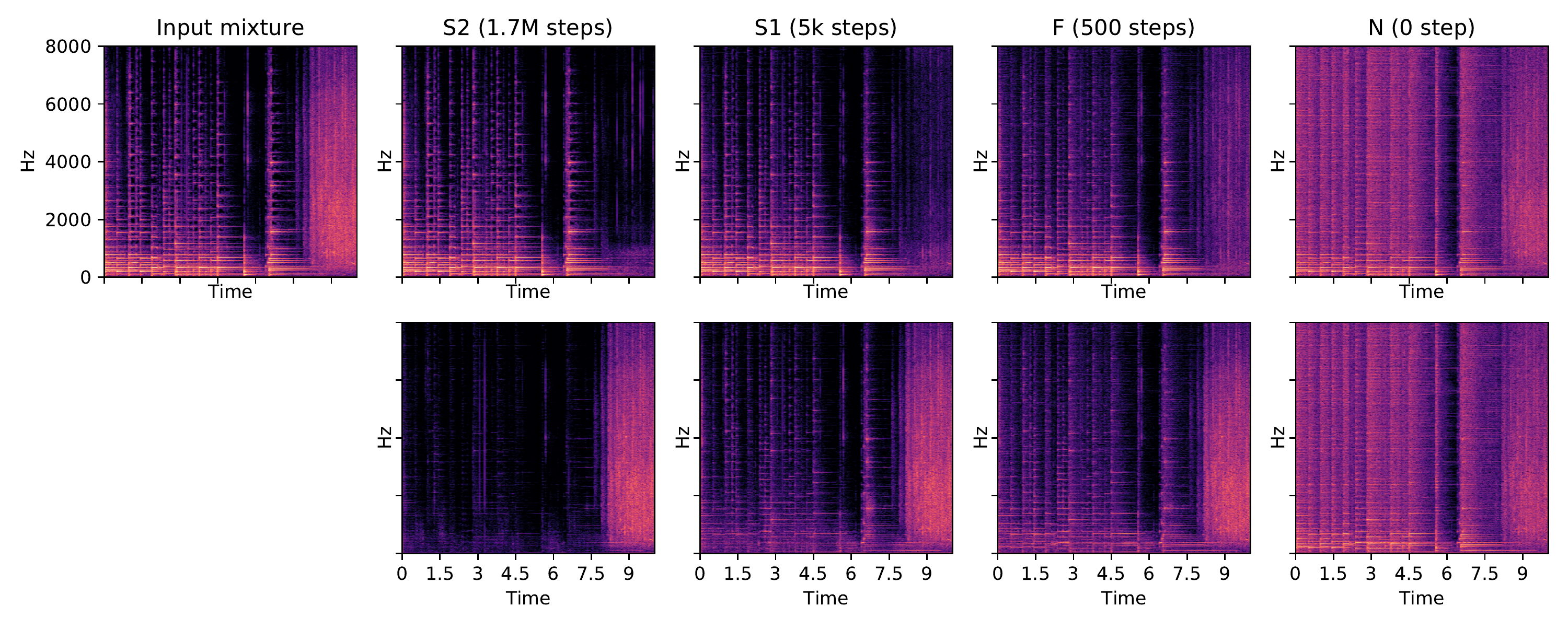}}
  \vspace{-4mm}
  \caption{Spectrograms of the two separated channels obtained with four checkpoints (S2, S1, F, N) of the same separation model, given one input mixture (top left). The input mixture contains a guitar melody (up to $\approx$8 s) followed by applause. For illustration purposes, this is a simple case where the separation is purely temporal (i.e., sources do not overlap). The general case features overlapping sources.}
  \label{fig:four_ckpts}
  \vspace{-2mm}
\end{figure*}
Table \ref{tab:mixit_ckpts} shows results of substituting S2 with the other identified processors, while keeping the rest of the framework as in Fig. \ref{fig:diagram}.
\begin{table}[!t]
\vspace{-2mm}
\caption{mAP using different checkpoints of the separation model as learning progresses (top), as well as some combinations (bottom). As back-end, the \textit{SimCLR} task is used (left), as well as the two proxy tasks trained jointly (right). TP and SpecAugment are applied. Comparison is always \textit{mix vs chan}.}
\vspace{+1mm}
\centering
\begin{tabular}{l|cc|cc}
\toprule
\textbf{Models}   & \multicolumn{2}{c|}{\textbf{SimCLR}}  & \multicolumn{2}{c}{\textbf{{SimCLR \& CP}}} \\
      & \textbf{QbE} & \textbf{Classif.} & \textbf{QbE} & \textbf{Classif.} \\
\midrule
\midrule
S2 (1.7M)         & 0.640 & 0.282                        & 0.649 & 0.289 \\
S1 (5k)           & 0.639 & 0.283                        & 0.651 & 0.293 \\
F (500)           & 0.651 & 0.280                         & 0.659 & 0.297 \\
N (0)             & \textbf{0.659} & \textbf{0.286}   & \textbf{0.663} & \textbf{0.301} \\
\midrule
S2 $\lor$ F 	    & 0.653 & 0.283                    & 0.658 & 0.300 \\
S2 $\lor$ N 	    & 0.660 & \textbf{0.297}             & 0.671 & 0.306 \\
S2 $\lor$ F $\lor$ N & \textbf{0.667} & 0.285          & \textbf{0.672} & \textbf{0.310} \\
\bottomrule
\end{tabular}
\label{tab:mixit_ckpts}
\vspace{-4mm}
\end{table}
By looking at the top left section of Table \ref{tab:mixit_ckpts}, two observations can be made:
First, the four processors all provide valid forms of augmentation to generate positive views for contrastive learning.
While sound separation is beneficial (S2), a poorer separation is also valuable (S1) and the earlier checkpoints of the separation network (which do not actually provide separation) are also useful (F and N).
Second, an untrained, quasi-random TDCN++ provides structured noise that surprisingly yields the best single-checkpoint performance (N).

Following the common practice of composing augmentations to achieve more powerful representations \cite{2020_ICML_SimCLR,fonseca2021unsupervised}, we investigate combining the processors.
The bottom left section of Table \ref{tab:mixit_ckpts} shows the best results obtained when combining processors using the \textit{OR} rule, that is, applying only one randomly selected processor at a time.
It can be seen that sound separation and the quasi-random TDCN++ noise turn out to be complementary augmentations, resulting in a more beneficial composition.
Adding the F processor seems to provide lift for QbE mAP, but not for classification mAP.
Applying two processors in cascade to every example does not improve performance.

\subsection{Joint Learning Framework}
\label{ssec:exp_joint}
Lastly, the right side of Table \ref{tab:mixit_ckpts} lists the results when training the entire framework of Fig. \ref{fig:diagram}, jointly optimizing both proxy tasks.
We observe trends similar to using the \textit{SimCLR} back-end alone (left side of Table \ref{tab:mixit_ckpts}), but with increased performance.
When compared to the \textit{CP} back-end alone (i.e., S2 in Table \ref{tab:mixit_ckpts} vs second row of Table \ref{tab:task_cp}), QbE mAP is improved by a large margin whereas classification mAP is on par.
Overall, while the boost from jointly optimizing both tasks is sometimes not very large, it is consistent across almost all cases considered, both for individual processors as well as their combinations.
We also note that the changes needed in the framework to accommodate a second task are minimal---only an additional MLP head and corresponding loss function---and the training setup carries no modifications---both tasks are trained jointly from scratch using one optimizer.
Adopting a curriculum learning instead could enhance performance \cite{jansen2020coincidence}.

Results suggest that the key ingredient is not the quality of the sound separation process, but rather the combination of diverse processing provided by the separation model as its learning progresses.
While training a separation model requires a certain effort, once it is done several non-parametric augmentation generators become available, facilitating the generation of useful positive examples.
While we choose a MixIT-based TDCN++, any source separation methodology could be used (and there may be additional benefit to using supervised systems).

\subsection{Comparison with Previous Work}
\label{ssec:exp_previous_work}
Table \ref{tab:sota} compares our best setup with previous work on the downstream classification task (see Sec. \ref{ssec:eval}).
Works are grouped by ascending embedding dimensionality, $d$.
\begin{table}[!t]
\vspace{-2mm}
\caption{Comparison with previous work using shallow model classification. mAP reported is classification mAP. MM = Multimodal approach.}
\vspace{+1mm}
\centering
\begin{tabular}{lccc}
\toprule
\textbf{Method}  & \textbf{$d$} & \textbf{MM} & \textbf{mAP} \\
\midrule
\midrule
Unsupervised triplet \cite{jansen2018unsupervised}  & 128   & -    & 0.244 \\
$C^3$ \cite{jansen2020coincidence}      & 128            & \checkmark        & 0.285 \\
Separation-based framework (ours)       & 128            & -      & 0.310    \\
\midrule
CPC  \cite{wang2020contrastive}         & 512            & -      & 0.277 \\
\midrule
Separation-based framework (ours)       & 1024           & -       & 0.326    \\
\midrule
MMV \cite{alayrac2020self}              & 2048           & \checkmark      & 0.309 \\
Multi-format \cite{wang2020multi}       & 2048           & -      & \textbf{0.329} \\
\midrule
$L^3$ \cite{arandjelovic2017look}       & 6144           & \checkmark    & 0.249 \\
\midrule
\textcolor{mygray}{Supervised PANN} \cite{kong2019panns} & - & - & \textcolor{mygray}{0.439}    \\
\textcolor{mygray}{Supervised PSLA}  \cite{gong2021psla} & -& -  & \textcolor{mygray}{0.474}    \\
\bottomrule
\end{tabular}
\label{tab:sota}
\vspace{-4mm}
\end{table}
Results are strictly comparable only in the top section as those works are the only ones using the same training data, evaluation protocol and downstream embedding dimensionality, $d=128$.
Note that $C^3$ is based on audio-video multimodality for representation learning \cite{jansen2020coincidence}, while our proposed framework outperforms it using only audio.
We also compare our system with works that use somewhat different evaluation settings in terms of, e.g., training data, embedding dimensionality or shallow classifier type, thus hindering a fair comparison.
For example, most previous works use larger $d$ values, ranging from 512 to 6144; we expect performance to improve to some extent as $d$ increases \cite{kong2019panns}---the embedding representation contains more information that can be leveraged by the shallow model in the downstream task.
We confirm this by increasing our $d$ from 128 to 1024, which yields an absolute increase of 0.016 mAP.
Some of these works leverage multimodal data for training such as audio-video ($L^3$ \cite{arandjelovic2017look}) or audio-video-text (MMV \cite{alayrac2020self}), while reporting worse performance than our lower-$d$ audio-only framework.
The current unsupervised state-of-the-art on this task is achieved by a contrastive learning setup that maximizes the agreement between raw audio and its spectral representation \cite{wang2020multi}.
Among several variants proposed by the authors, we select the one that is more comparable to our proposed framework (i.e., using only log-mel as input and one encoder).
Our reported performance is on par with this approach (0.326 vs 0.329) despite it leveraging higher $d$ (2048 vs our 1024)
and a much larger batch size (32768 vs our 64), potentially having an impact on performance as discussed in Sec. \ref{ssec:imp_details}.
Better results are reported in previous work \cite{wang2020multi} by combining two different encoders (one per audio format) and concatenating their output representations. 
Finally, for reference, we include the current supervised state-of-the-art on this task.
PANN is based on data balancing and augmentation \cite{kong2019panns}, whereas PSLA makes use of a collection of training techniques to boost performance (ImageNet pretraining, data balancing and augmentation, label enhancement and model aggregation) \cite{gong2021psla}.


\section{Conclusion}
We have presented a sound separation-based contrastive learning framework for unsupervised audio representation learning. 
We show that sound separation can be seen as a valid augmentation to generate positive views for contrastive learning, and that learning to associate sound mixtures with their constituent separated channels elicits semantic structure in the learned representation, outperforming comparable systems without separation.
We demonstrate that sound separation can be successfully combined with other commonly-used augmentations to define more challenging proxy tasks.
We discover that the transformations provided by different checkpoints of the same separation model as learning progresses are valid, and sometimes complementary, forms of augmentation for generating positives.
In addition, we show the benefit of jointly training the proxy tasks of similarity maximization and coincidence prediction.
By appropriately combining several separation processors followed by the joint optimization of the two proxy tasks, we obtain downstream AudioSet classification results competitive with the state-of-the-art in unsupervised representations, and outperforming multimodal approaches.  

\section{ACKNOWLEDGMENT}
The authors would like to thank Malcolm Slaney (Google Research) and Luyu Wang (DeepMind) for helpful feedback and discussions.

\bibliographystyle{IEEEtran}
\bibliography{refs21}
%
%
%
%
%
%
%
%
%

\appendix
\section{Selection of Number of Separated Channels}
\label{sec:selectM}
A critical parameter in the proposed framework is the number of output waveforms in the separation model, $M$, which must be defined at train time.
Upon inspection of a few AudioSet clips selected randomly, we realize that many clips contain one or two dominant sources (i.e., in the foreground, lasting long within the clip), sometimes accompanied by additional sources (either in the foreground but very short, e.g., impact sounds, or in the background).\footnote{The main exception to this rule is music segments.}
We therefore ran preliminary experiments with $M=\left\{ 2,4\right\}$ and saw that results using $M=4$ were slightly worse for both proxy tasks.
We attribute this to the fact that when $M=4$ it is not uncommon to find output channels which are almost empty, filled with mild background noise, or with sound sources active only in a very short period of time. 
We hypothesize that using these channels to create positive pairs can hurt performance.

To confirm our hypothesis, we designed simple heuristics (based on energy and cosine similarity) to detect these quasi-empty channels, in order to allow discarding the “worst” channel in every contrastive setup, thus keeping only the other 3 channels from where to pool positive examples.
This led to a small but consistent performance improvement, confirming our initial hypothesis, yet still underperforming results with $M=2$. 
While further optimizations to allow using $M=4$ could be pursued, for simplicity we decided to adopt $M=2$ for our experiments, which is the minimal separation possible. 
In some cases, the two output waveforms coming out of the separation model will contain one source each, whereas in other cases they will contain several sources each. 
Consequently, we use the term separated \textit{channels} (and not \textit{sources}) as it is deemed more appropriate.
We believe $M=2$ is sufficient to evaluate our hypothesis of sound separation serving as a valid transformation for view generation in contrastive learning.

\end{sloppy}
\end{document}